\documentclass[aps,superscriptaddress,twocolumn]{revtex4}

\renewcommand*{\[}{\begin{equation}}
\renewcommand*{\]}{\end{equation}}

\usepackage{multirow}
\usepackage{array}
\usepackage{graphicx}
\usepackage{amssymb}
\usepackage{epstopdf}
\usepackage{subfigure}
\usepackage{graphicx}
\usepackage{physics}

\usepackage{hyperref}
\usepackage{color}
\usepackage{float}

\begin{document}

\title{Extension of all-optical reconstruction method for isolated attosecond pulses using high-harmonic generation streaking spectra}

\author{Kan Wang}
\author{Yong Fu}
\author{Baochang Li}
\author{Xiangyu Tang}
\author{Zhong Guan}
\author{Bincheng Wang}
\affiliation{Department of Applied Physics, Nanjing University of Science and Technology, Nanjing, Jiangsu 210094, China}

\author{C. D. Lin}
\affiliation{J. R. Macdonald Laboratory, Department of Physics, Kansas State University, Manhattan, Kansas 66506, USA}

\author{Cheng Jin}
\thanks{Corresponding author. E-mail: cjin@njust.edu.cn}
\affiliation{Department of Applied Physics, Nanjing University of Science and Technology, Nanjing, Jiangsu 210094, China}
\affiliation{MIIT Key Laboratory of Semiconductor Microstructure and Quantum Sensing, Nanjing University of Science and Technology, Nanjing, Jiangsu 210094, China}

\date{\today}

\begin{abstract}
An all-optical method for directly reconstructing the spectral phase of isolated attosecond pulse (IAP) has been proposed recently [New J. Phys. 25, 083003 (2023)]. This method is based on the high-harmonic generation (HHG) streaking spectra generated by an IAP and a time-delayed intense infrared (IR) laser, which can be accurately simulated by an extended quantitative rescattering model. Here we extend the retrieval algorithm in this method to successfully retrieve the spectral phase of an shaped IAP, which has a spectral minimum, a phase jump about $\pi$, and a ``split" temporal profile. We then reconstruct the carrier-envelope phase of IR laser from HHG streaking spectra. And we finally discuss the retrieval of the phase of high harmonics by the intense IR laser alone using the Fourier transform of HHG streaking spectra.
\end{abstract}

\maketitle
\section{Introduction}

%
%

The revolution of attosecond science has enabled to investigate and manipulate the ultrafast electronic dynamics in atoms, molecules, and solids with an unprecedented temporal resolution \cite{Krausz-rmp-2009,Peng-pr-2015,calegari-jpb-2016,yun-jpb-2017}. As its foundation, the technique of high-harmonic generation (HHG) from the gas medium has become mature and stable to efficiently produce attosecond pulse trains \cite{Agostini-sci-2001} or isolated attosecond pulses (IAPs) \cite{krausz-nat-2001}. Especially, with the development of mid-infrared laser and phase-locking technologies, IAPs with ultrashort temporal duration down to a few tens of attoseconds can be obtained up to the water-window spectral region \cite{jieli-nc-2017,Gaumnit-oe-2017}. To fully characterize an IAP, the so-called attosecond streaking method is commonly used, in which the IAP first ionizes an atomic target and then a moderate infrared (IR) laser is applied to modulate the photoelectron spectra \cite{krausz-prl-2010}.  A number of algorithms have been developed to retrieve the IAP and the IR laser from time-delay dependent photoelectron spectra, including the FROG-CRAB (frequency-resolved optical gating for complete reconstruction of attosecond bursts) \cite{mairesse-pra-2005}, PROOF (phase retrieval by omega oscillation filtering) \cite{zchang-oe-2010}, PROBP (phase retrieval of broadband pulses) \cite{zhaox-pra-2017,zhaox-prall-2020}, etc.

Alternatively, the IAP can be combined with a time-delayed, few-cycle intense IR laser to perturb or reform the HHG process by the IR laser alone. The modulated harmonic spectra with the time delay between the IAP and the IR laser is called the high-harmonic generation streaking spectra, which provides with an all-optical scheme to retrieve the IAP \cite{kanw-pra-2021}. The IAP in the extreme ultraviolet (XUV) can actually affect the well-known three steps in the HHG process differently. For instance, the XUV pulse can populate electrons to excited states through the single-photon ionization, leading to an increase in the ionization energy, which in turn affects the efficiency and cutoff energy of harmonic emission \cite{acbrown-prl-2016}. The IAP can act in the propagation step and cause multiple scatterings of active electrons, thereby influencing the low-energy high harmonics \cite{lewenstein-pra-1994}. If the electrons absorb high-energy XUV photons in the returning step, it could broaden the HHG plateau region \cite{Sarantseva-pra-2020}. Furthermore, XUV pulses can be employed to selectively control the specific electron trajectory for modulating the harmonic field \cite{zhang-pra-2009}. The XUV parametric amplification (XPA) has been also achieved when the XUV pulse was combined with the intense IR laser, attributing to the nonadiabtic forward scattering of electron wavepacket \cite{serrat-prl-2013}. To precisely simulate the HHG streaking spectra under the interplay of the IAP and the time-delayed IR laser, we developed an extended quantitative rescattering (EQRS) model \cite{kanw-pra-2021} by greatly improving the strong-field approximation (SFA) \cite{lewenstein-pra-1994,serrat-oe-2015,Serrat-oe-2016}. Based on the EQRS model and the SFA, we derived a retrieval algorithm for successfully reconstructing the IAPs both in the XUV and soft X-ray regions from the HHG streaking spectra \cite{kanw-njp-2023}. However, there are still some challenging questions not resolved yet. Can one retrieve the information of IR laser from the HHG streaking spectra, for example, the carrier-envelope phase (CEP) of IR laser? It is known that the reconstruction of IR laser is feasible in the traditional method of attosecond streaking camera \cite{zhaox-pra-2017}. Can one obtain the information of HHG by the intense IR laser alone?

It has been established that the minimum structure existing in the atomic HHG spectra can be directly related to that in the photorecombination cross-section (PRCS) \cite{qrs-pra-2009}. A typical example is the Cooper minimum (CM) in the HHG spectra of Ar \cite{sakai-pra-2008,worner-prl-2009,schafer-pra-2011,higuet-pra-2011}. By spectral filtering the high harmonics around the CM, the temporal XUV pulse would be severely shaped \cite{Schoun-prl-2014}. The CM in the HHG spectrum has also been observed in various atoms \cite{shiner-jpb-2012} and molecules \cite{bruner-fd-2016,suarez-pra-2017,ruberti-pccp-2018}. Recently, Jin \textit{et al.} \cite{jinc-pra-2020,jin-pra-2020} reported on the control of minimum in the HHG spectra of aligned CO$_{2}$ molecules by adjusting the degree of alignment. Such minimum is similar to the CM, called as the ``structural minimum", can be explained by the alignment averaged PRCS. Since the amplitude of averaged PRCS varies rapidly and its phase changes near $\pi$ around the minimum, the resulted temporal attosecond pulses are greatly shaped and can be tuned. Can such structured attosecond pulses also be reconstructed from the HHG streaking spectra?

To answer above questions, in this paper, we will extend and improve the all-optical retrieval algorithm based on the HHG streaking spectra. We will specifically target on how to correctly retrieve the spectral phase of a shaped isolated attosecond pulse in Sec. \ref{s2}, how to identify the CEP of IR laser from HHG streaking spectra in Sec. \ref{s3}, and how to obtain the phase of high harmonics by the IR laser alone in Sec. \ref{s4}. Finally, the conclusions will be given in Sec. \ref{s5}.

\section{Reconstruction of spectral phase of shaped attosecond pulses}\label{s2}
\begin{figure}[htbp]
\centering\includegraphics[scale=0.27]{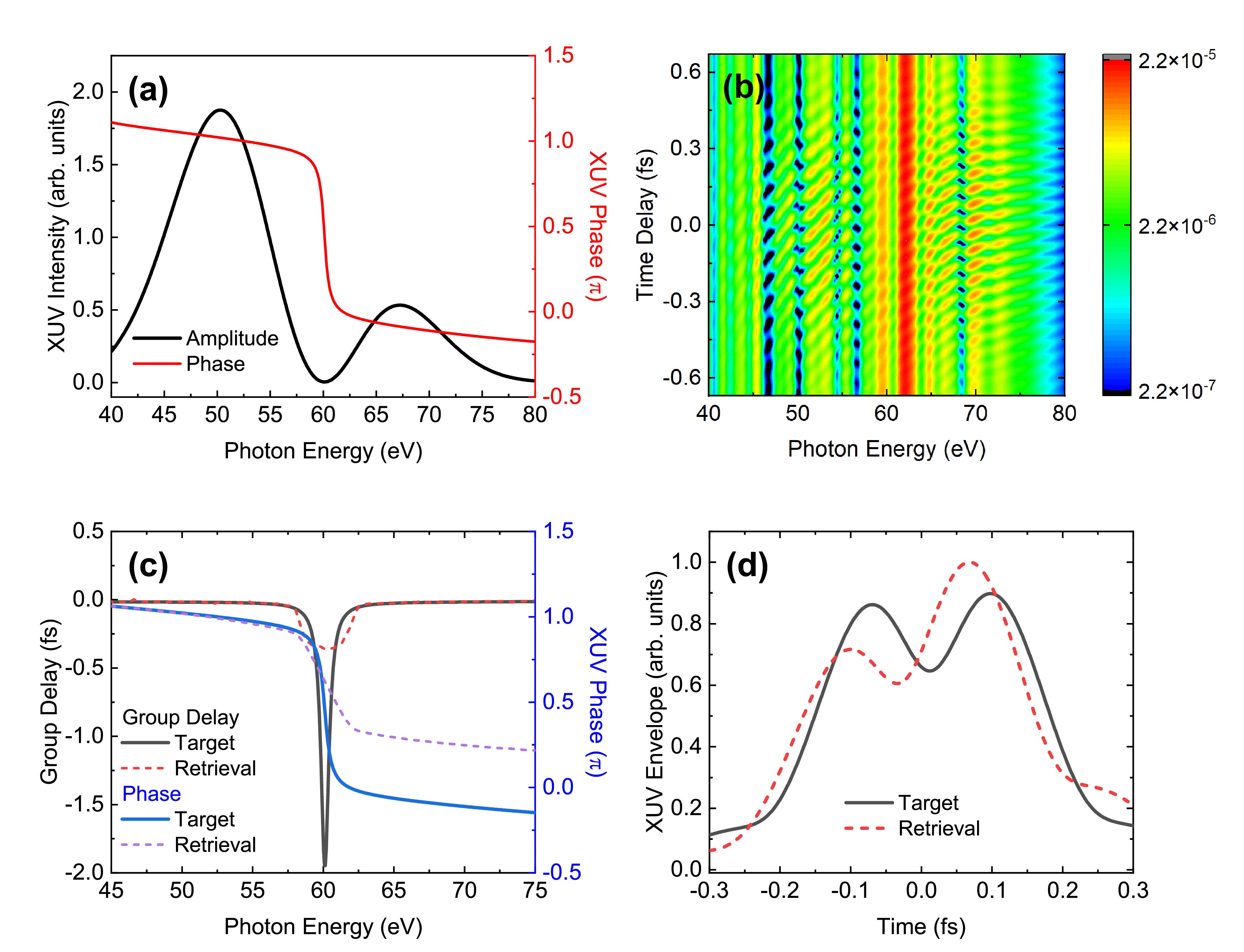}
\caption{(a) The spectral amplitude and phase of shaped attosecond pulse. (b) HHG streaking spectra calculated by the EQRS model. (c) The group delay and spectral phase obtained by the original phase reconstruction method in comparison with input values. (d) Comparison of reconstructed temporal profile of shaped IAP with the input one. \label{fig-1}}
\end{figure}

First, we simulate the HHG streaking spectra of Ne atom under a shaped IAP and a time-delayed IR laser. The shaped IAP exhibits an extreme minimum centered at 60 eV in the spectral amplitude and a significant phase jump as shown in Fig. \ref{fig-1}(a). These data are taken from Ref. \cite{jinc-pra-2020} when the HHG of aligned CO$_{2}$ molecules is generated under the alignment degree of 0.40 and the pump-probe angle of $0^{\circ}$. In the simulations, we use an IR laser with the wavelength of 800 nm, the full-width-at-half-maximum (FWHM) duration of 5 fs, the peak intensity of 2.5 $\times$ 10$^{14}$ W/cm$^2$, and the CEP of 0. We set the peak intensity of the shaped IAP as 2.5 $\times$ 10$^{11}$ W/cm$^2$, and calculate the HHG streaking spectra with the time delay between the IAP and IR laser by using the EQRS model. The results are shown in Fig. \ref{fig-1}(b). The modulation structure can be clearly observed with the time delay if the spectral amplitudes of shaped IAP are stronger, for example, around 50 eV and 67 eV, but it is difficult to be distinguished at the minimum energy, i.e., at 60 eV.

We then apply the all-optical phase reconstruction method developed in Ref. \cite{kanw-njp-2023}. Note that we focus on the retrieval of spectral phase of shaped IAP only while the spectral intensity can be easily measured by the spectrometer, which is assumed known. At each photon energy, by fitting the time-delay modulated spectrum, we obtain the time delay corresponding to the maximum modulation, and construct the curve of group delay, as shown by the red dashed line in Fig. \ref{fig-1}(c). The reconstructed curve deviates much from the input one (black solid line) around the minimum. This leads to some big errors in the reconstructed phase of shaped IAP (purple dashed line) compared to the input one (blue solid line). We show the retrieved XUV pulse in time (red dashed line) in Fig. \ref{fig-1}(d), and the input pulse (black solid line) is also plotted for comparison. The retrieval procedure returns errors in the peak positions, relative peak intensities, and widths of temporal shaped pulse. Such errors cannot be acceptable for characterizing the shaped IAP.

\begin{figure}[htbp]
\centering\includegraphics[scale=0.27]{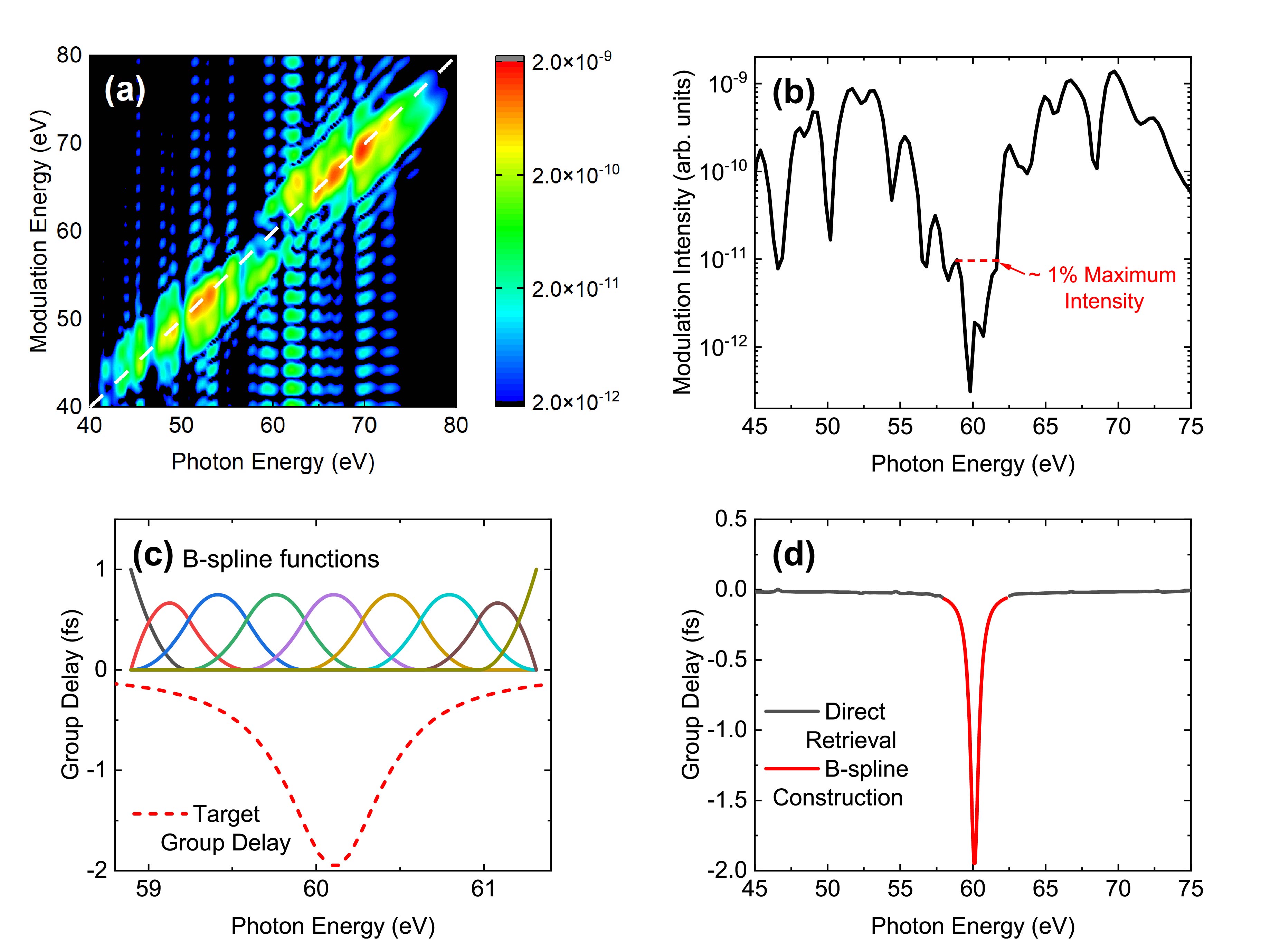}
\caption{(a) The modulation energy spectra obtained by performing the Fourier transformation with respect to the time-delay. (b) The intensity of modulation spectrum with the photon energy along the diagonal line. (c) The B-spline basis functions and the target group delay of shaped IAP. (d) Reconstructed group delay contains two parts: one is from the original phase reconstruction method and the other is newly constructed by the B-spline functions.  \label{fig-2}}
\end{figure}

To explain the errors occurring in the extraction of group delay, we make a transformation of HHG streaking spectra in Fig. \ref{fig-1}(b). For each photon energy $\omega$, the Fourier transform with respect to the time delay is performed, which can be expressed as
\begin{equation}
\label{tdinfouriertrans}
S'(\omega,\omega_{mod})=\frac{1}{2\pi}\int_{\tau_{1}}^{\tau_{2}}S(\omega,\tau)e^{i\omega_{mod}\tau}d\tau,
\end{equation}
where $S(\omega,\tau)$ represents the HHG streaking spectra and $S'(\omega,\omega_{mod})$ is the modulated energy spectrum after the Fourier transform. $\omega_{mod}$ is the modulation frequency, reflecting the frequency components of the modulation structure, and $\tau_{1}$ and $\tau_{2}$ are the start and end time delays for the integral, respectively. The modulated energy spectra are shown in Fig. \ref{fig-2}(a). One can see that the considerable modulation intensities are primarily distributed along a diagonal line for most photon energies. This indicates that the major modulation frequency in the HHG streaking spectra is about the same as the spectral frequency of IAP. However, for photon energies around 60 eV, the peak of modulation intensity is split along the modulation energy, due to the interference between the adjacent low-energy and high-energy modulations. This also accounts for errors of extracting the group delay by using the phase reconstruction method.

\begin{figure}[htbp]
\centering\includegraphics[scale=0.27]{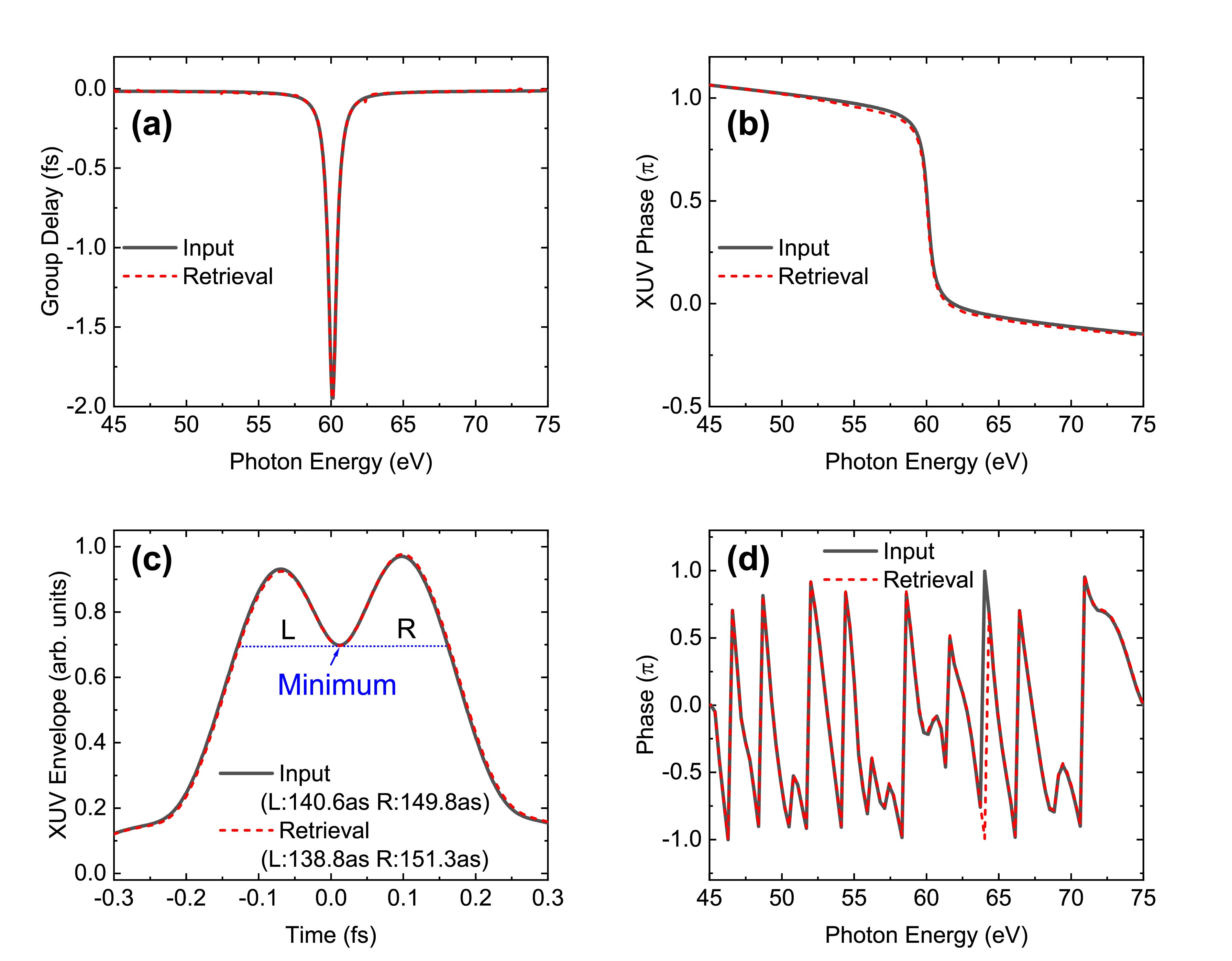}
\caption{The comparison of retrieved results with input ones: (a) the group delay, (b) the spectral phase, and (c) the temporal profile of shaped IAP, and (d) the phase of the modulation energy spectra. \label{fig-3}}
\end{figure}

We next improve the phase reconstruction method, specifically, eliminating the errors in the reconstructed group delay. From Fig. \ref{fig-1}(c), one can see that the reconstructed results only need to be fixed in a narrow energy region around the spectral minimum of the IAP. We first determine the range of such photon-energy region. In Fig. \ref{fig-2}(b), the modulation intensity along the diagonal line in Fig. \ref{fig-2}(a) is plotted as a function of photon energy. By testing various shaped IAPs, a photon-energy region is selected where the modulation intensities are within 1 \% of the maximum value (red dashed line) as shown in Fig. \ref{fig-2}(b). We next determine how to construct the group delay in this energy region, which has a sharp structure. We adopt the B-spline functions as a basis to expand the group delay in the following:
\begin{equation}
\label{B-group}
f_{gd}(\omega)=\sum_{i=1}^{n}g_{i}B_{i}^{k}(\omega),
\end{equation}
where $g_{i}$ is the expansion factor for the $i$-th B-spline function with the $k$-th order. It is defined as
\begin{equation}
B_{i}^{k}(\omega)=\left\{\begin{matrix} 1,
 & \omega_{i}\leq\omega\leq\omega_{i+1}\\ 0,
 & otherwise,
\end{matrix}\right.
\end{equation}
\begin{equation}
B_{i}^{k}(\omega)=\frac{\omega-\omega_{i}}{\omega_{i+k-1}-\omega_{i}}B_{i}^{k-1}(\omega)+\frac{\omega_{i+k}-\omega}{\omega_{i+k}-\omega_{i+1}}B_{i+1}^{k-1}(\omega).
\end{equation}
Here ${\omega_i}$ is the frequency knot points. Details of B-spline functions can be founded in Refs. \cite{zhaox-pra-2017,Jin-ctp-2006}. Fig. \ref{fig-2}(c) illustrates the B-spline basis functions and the target group delay. Here we utilize nine B-spline functions with the 9-th order. Thus 9 coefficients $g_{i}$ in Eq. (\ref{B-group}) need to be determined. The resulted group delay in the narrow energy region can be combined with that obtained by the previous phase construction method as shown in Fig. \ref{fig-2}(d). We also need to determine another parameter $\beta$, connecting the phase over the entire spectral region. Then, we employ the genetic algorithm (GA) to optimize the coefficients $g_{i}$ and the parameter $\beta$. In each iteration, by combing the spectral amplitude and the phase given by $\{g_{i}, \beta\}$, we obtain the shaped IAP in time, calculate the HHG streaking spectra by using the EQRS model, and simulate the modulation energy spectra according to Eq. (\ref{tdinfouriertrans}). The population size is chosen to be 5 and the maximum number of generation is set to be 1000 - 2000 to guarantee the convergence. Finally, we give the error (or fitness) function in the optimization. We choose the phase of modulation spectra $\phi_{mod}(\omega)$ as a function of photon energy along the diagonal line in Fig. \ref{fig-2}(a). This phase is plotted in Fig. \ref{fig-3}(d). The error function can be expressed as
\begin{equation}
\label{errorfunc}
F_{\text{error}}(g_{i}, \beta) = \frac{1}{m}\sum_{m} |\phi_{mod,0}(\omega_{m})-\phi_{mod,1}(\omega_{m})|,
\end{equation}
where $\phi_{mod,0}(\omega_{m})$ and $\phi_{mod,1}(\omega_{m})$ represent the phase of input and reconstructed modulated energy spectrum, respectively. The photon energy sampling points $\omega_m$ in the whole spectral region are chosen as $m$ = 100.

\begin{figure*}[t]
\centering\includegraphics[scale=0.33]{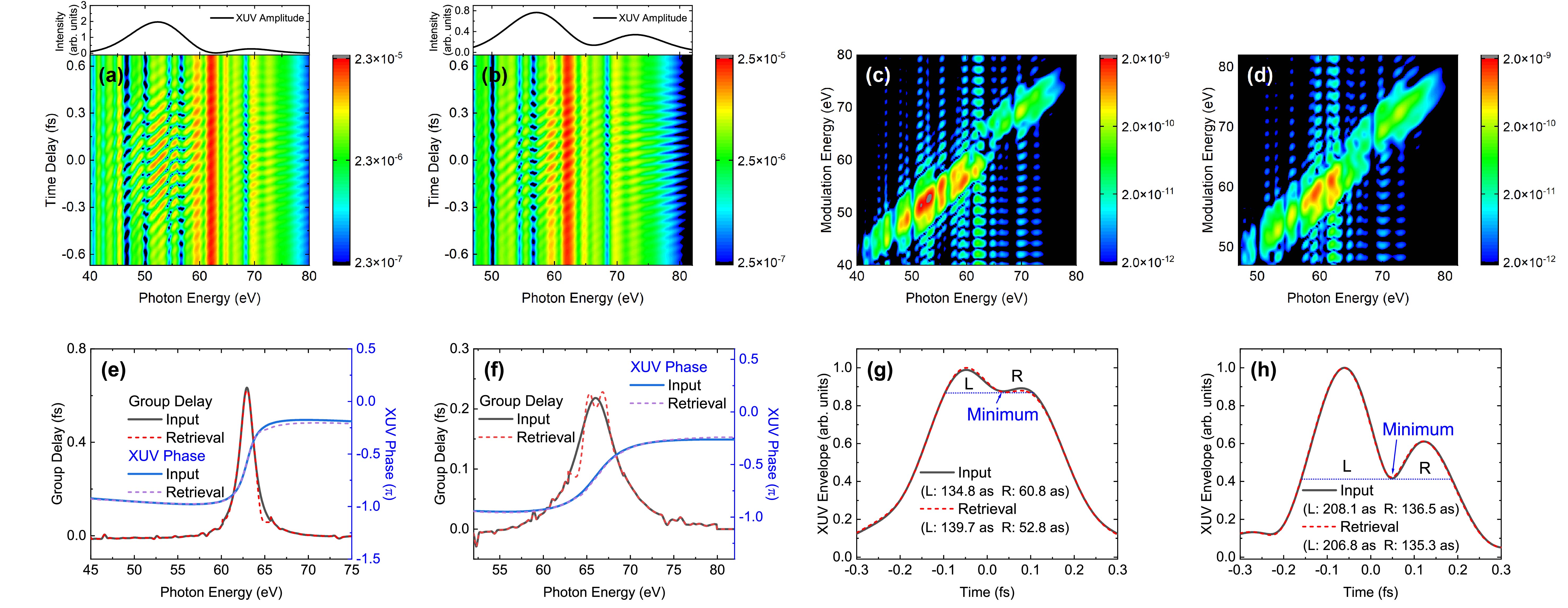}
\caption{(a, b) HHG streaking spectra and (c, d) modulation energy spectra through the Fourier transform with the time delay by using two shaped IAPs, respectively. The spectral amplitudes of two IAPs are plotted on the top in (a, b). Comparisons of retrieved results of two shaped IAPs with input ones: (e, f) the group delay and spectral phase, and (g, h) the temporal profile.  \label{fig-4}}
\end{figure*}
For the shaped IAP displayed in Fig. \ref{fig-1}, we follow the above steps and obtain the reconstructed results in Fig. \ref{fig-3}. Comparisons of the input and reconstructed group delay and spectral phase of the IAP are present in Figs. \ref{fig-3}(a) and (b), respectively. The corresponding temporal profiles are shown in Fig. \ref{fig-3}(c). We use a different way to define the width of double peaked pulse. We identify the position of minimum in the temporal profile, use its intensity as a reference, and define the widths of ``left" and ``right" peaks separately. Compared to the input information (i.e., the width of ``left" peak is 140.6 as and the ``right" one is 149.8 as), the reconstructed widths are 138.8 as and 151.3 as for ``left" and ``right" peak, respectively, which have an error of less than 2 \%. We also compare of input and reconstructed phases of modulation spectra in Fig. \ref{fig-3}(d). The dramatic change in the spectral phase around the minimum can be reflected in the phase of modulation spectra. Choosing the phase in the fitness function greatly improves the accuracy of retrieval procedure.

We then choose another two shaped IAPs with different temporal shapes and spectral phases around the minimum and test the validity of our improved reconstruction method. As demonstrated by Jin \textit{et al}. \cite{jinc-pra-2020}, such shaped pulses can be obtained by tuning the alignment degree of CO$_2$ molecules or by adjusting the pump-probe angle. Simulated HHG streaking spectra by using the EQRS model are shown in Figs. \ref{fig-4}(a) and (b), with the corresponding spectral amplitude distributions being plotted on the top. We apply the Fourier transform with respect to the time delay and obtain the modulation energy spectra in Figs. \ref{fig-4}(c) and (d). These spectra exhibit the typical interference patterns and the splitting around the spectral minima. The results of group delay and spectral returned by the improved reconstruction method are shown in Figs. \ref{fig-4}(e) and (f), respectively. Although there are some small deviations in the reconstructed results of group delay, the spectral phases can achieve the good agreement with the input ones. The temporal profiles of two reconstructed pulses are plotted in Figs. \ref{fig-4}(g) and (h). For the fist IAP, the retrieved width of ``left" peak is 139.7 as compared to the input value of 134.8 as, and it is 52.8 as for the ``right" peak while the input width is 60.8 as. The error in the retrieved width is increased for the ``right" peak because its peak intensity is in close proximity to the minimum one.  For the second IAP, the reconstructed width of ``left" peak is 206.8 as (compared to the input value of 208.1 as) and the ``right" one is 135.3 as (compared to the input value of 136.5 as), with the error less than 1 \%.

\section{Reconstruction of CEP of IR pulse from HHG streaking spectra}\label{s3}
We then discuss if the CEP of few-cycle intense IR laser is imprinted in the HHG streaking spectra. In Ref. \cite{kanw-njp-2023}, we derived that the coupling term of the IAP and the IR laser in the induced dipole moment can be written as 
\begin{equation}
\label{proptoeqx2}
|x_{2}(\omega,\tau)|^{2} \propto E_{\text{IR}}^{2}(\alpha-\tau),
\end{equation}
where $\alpha$ is the derivative of the XUV spectral phase with the time. In this equation, the CEP $\Phi_{\text{CEP}}$ of the IR laser is assumed as 0. When it becomes arbitrary, Eq. (\ref{proptoeqx2}) can be extended as
\begin{equation}
\label{proptoeqx3}
|x_{2}(\omega,\tau)|^{2} \propto E_{\text{IR}}^{2}(\alpha - \tau + \Phi_{\text{CEP}}/\omega_{\text{IR}}),
\end{equation}
where $\omega_{\text{IR}}$ is the angular frequency of the IR laser. Thus, the time delay corresponding to the maximum modulation can be written as
\begin{equation}
\label{tauomega}
\tau(\omega) = \alpha + \frac{\Phi_{\text{CEP}}}{\omega_{\text{IR}}}.
\end{equation}
Eq. (\ref{tauomega}) means that the time delay corresponding to the maximum modulation intensity in the HHG streaking spectra is varied with the CEP of the IR laser.

\begin{figure}[htbp]
\centering\includegraphics[scale=0.33]{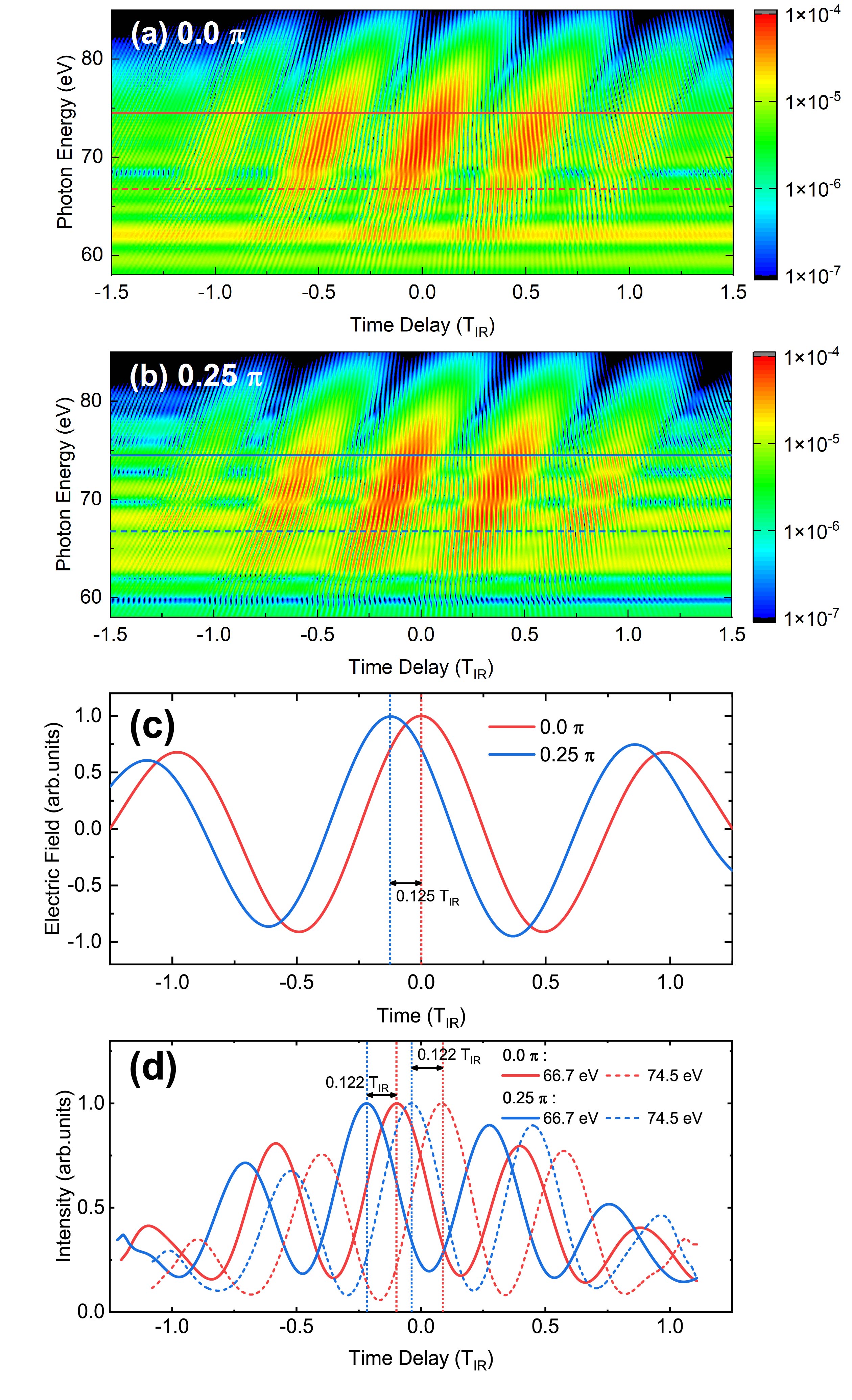}
\caption{High-harmonic streaking spectra generated by the IR laser with different CEP: (a) 0.0 $\pi$ and (b) 0.25 $\pi$. The corresponding electric fields of the IR laser are shown in (c). (d) Envelopes of time-delayed HHG modulation intensity for different combinations of the CEP and photon energy: 0.0 $\pi$, 66.7 eV (red solid line); 0.0 $\pi$, 74.5 eV (red dashed line); 0.25 $\pi$, 66.7 eV (blue solid line); and 0.25 $\pi$, 74.5 eV (blue dashed line). \label{fig-5}}
\end{figure}

To test it, we choose a regular chirped IAP centered at 71.3 eV with a spectral bandwidth of 9 eV, and set the CEP of the IR laser as 0 and 0.25 $\pi$, respectively. The resulted HHG streaking spectra by using the EQRS model are shown in Figs. \ref{fig-5}(a) and (b). For both cases, the modulation intensity along the time delay follows the periodicity of half optical cycle of the IR laser. With the change of the CEP, the overall modulation structure undergoes a shift in time delay. We choose photon energies of 66.7 (solid lines) and 74.5 eV (dashed lines) and extract the data of modulated HHG intensity as a function of time delay. After fitting these data, the obtained oscillating envelopes are plotted in Fig. \ref{fig-5}(d). Vertical dotted lines are used to mark the position of peak in each curve. For both photon energies, the extracted shift in time delay due to the change of the IR CEP is 0.122 $T_{\text{IR}}$, where $T_{\text{IR}}$ is the period of the IR laser. Compared to the electric fields of the 5 fs IR laser plotted in Fig. \ref{fig-5}(c), the time shift is 0.125 $T_{\text{IR}}$, which accounts for the change of electric peak due to the change of the CEP. The reconstructed time delay or the CEP equivalently according to Eq. (\ref{tauomega}) has an error of 2.4 \%.

\section{Reconstruction of phase of high harmonics driven by IR laser alone}\label{s4}
According to the EQRS model, the induced dipole moment $x(\omega,\tau)$ under the IAP and the time-delayed IR laser is mostly from two terms \cite{kanw-njp-2023}. The first one is the induced dipole moment by the IR laser alone, which can be expressed as
\begin{equation}
x_{1}(\omega)=F(\omega)e^{i\phi_{1}(\omega)}.
\end{equation}
Here $F(\omega)$ is the spectral amplitude and $\phi_{1}(\omega)$ is the spectral phase of $x_{1}(\omega)$. The second one is induced dipole moment due to the coupling of the IAP and the IR laser, and can be written as
\begin{equation}
x_{2}(\omega,\tau)\propto \frac{\epsilon}{2}e^{i(\phi_{\text{XUV}}(\omega)-\omega\tau)}E_{\text{IR}}(\alpha-\tau),
\end{equation}
where $\phi_{\text{XUV}}(\omega)$ is the spectral phase of the IAP. We assume that the amplitudes of two terms are comparable, and only keep the interference part. The intensity of HHG streaking spectra, i.e., $|x(\omega,\tau)|^2$, can be simplified as
\begin{equation}
\label{smod}
S(\omega,\tau) \approx A^2 \cos[\phi_{1}(\omega)-\phi_{\text{XUV}}(\omega)+\omega \tau].
\end{equation}
We then perform the Fourier transform with respect to the time delay according to Eq. (\ref{tdinfouriertrans}). And the phase of the modulated energy spectra is then written as \cite{dong-oe-2020}
\begin{equation}
\label{phaserelation}
\phi_{mod}(\omega)=\phi_{1}(\omega)-\phi_{\text{XUV}}(\omega).
\end{equation}
Therefore, once the spectral phase of the IAP is correctly retrieved by the original (or improved) phase reconstruction method, the phase of HHG by the IR laser alone can also be retrieved from the HHG streaking spectra.
\begin{figure}[htbp]
\centering\includegraphics[scale=0.23]{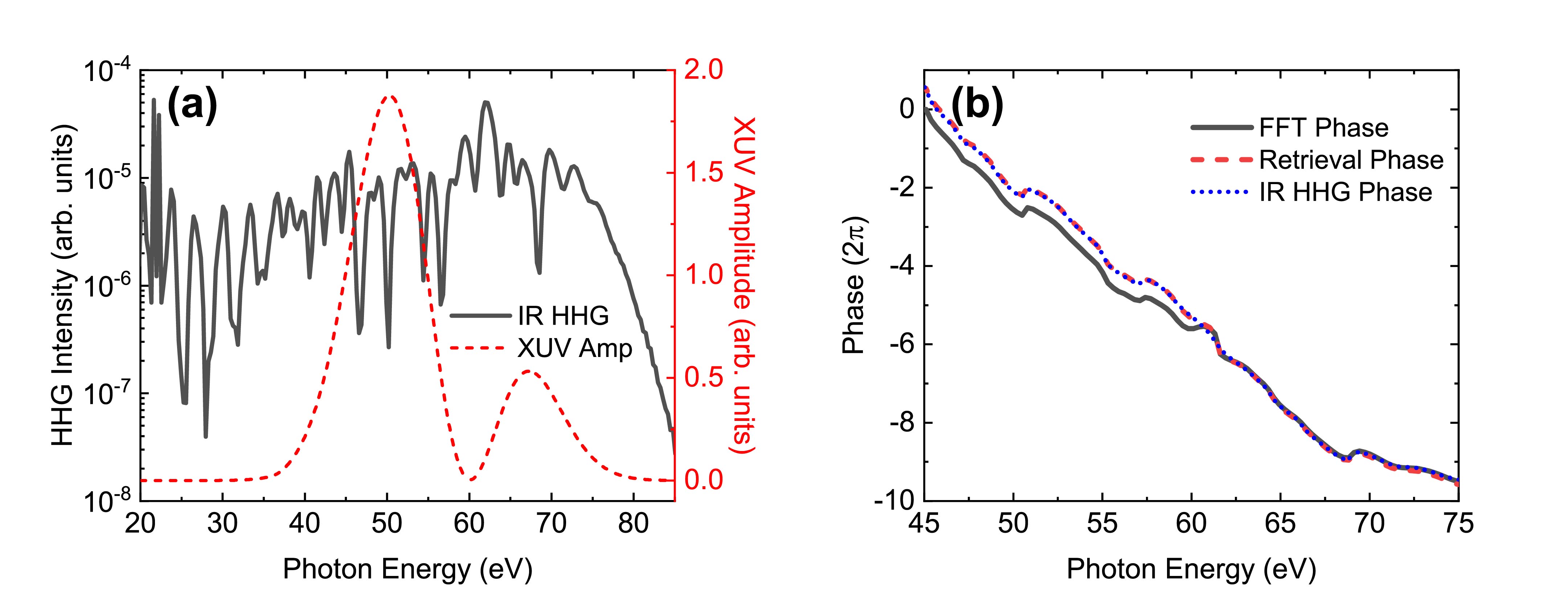}
\caption{(a) HHG spectra (black solid line) by the IR laser alone and the spectral amplitude (red dashed line) of shaped IAP. (b) The phase of the modulation energy spectra (black solid line), and the comparison of the retrieved HHG phase $\phi_{1}(\omega)$ by the IR laser alone and numerically calculated one by the QRS model. \label{fig-6}}
\end{figure}

We illustrate such an example in Fig. \ref{fig-6}. Fig. \ref{fig-6}(a) shows the HHG spectra (black solid line) driven by the single IR laser and the spectral amplitude distribution (red dashed line) of the shaped IAP. Note that the phase information can only be obtained for high harmonics within the spectral region of the shaped IAP. Fig. \ref{fig-6}(b) shows the phase (black solid line) of modulated energy spectra, which is obtained by smoothly connecting the phase along the photon energy in Fig. \ref{fig-3}(d). According to Eq. (\ref{phaserelation}), by subtracting the phase of the shaped XUV pulse as shown in Fig. \ref{fig-3}(b), the reconstructed phase $\phi_{1}(\omega)$ (red dashed line) can be obtained. It achieves a perfect agreement with the phase (blue dotted line) of single-IR HHG numerically calculated by the QRS model.

\section{Summary}\label{s5}
In summary, we modified the previously proposed all-optical phase reconstruction method by using high-harmonic generation (HHG) streaking spectra \cite{kanw-njp-2023}. We were able to successfully retrieve the spectral phase of shaped IAP, the CEP of IR laser, and the phase of high harmonics by the IR laser alone. First, we used B-spine basis functions, employed the genetic algorithm, and defined the fitness function in terms of the phase of modulation energy spectra. And it was combined with the previous phase reconstruction method to retrieve the spectral phase and the ``split" temporal information of shaped IAPs. Second, we included the CEP of IR laser in the formulation for the reconstructed XUV phase, it thus can be extracted from the HHG streaking spectra. Third, we related the phase of modulation energy spectra to the spectral phase of IAP and the phase of high harmonics by the IR laser alone. We showed that the phase of HHG by the single-IR laser can be precisely reconstructed if the spectral phase of IAP is known (or accurately retrieved). This work greatly extends the applicability of all-optical method based on the HHG streaking spectra.

\section{Acknowledgement}
This work was supported by National Natural Science Foundation of China (NSFC) under Grants Nos. 12274230, 12204238, and 11834004, Funding of Nanjing University of Science and Technology (NJUST) under Grant No. TSXK2022D005, and Natural Science Foundation of Jiangsu Province under Grant No. BK20220925. C.D.L. was supported by Chemical Science Division, Office of Basic Energy Sciences, Office of Science, US Department of Energy under Grant No. DE-FG02-86ER13491.

\end{document}